\documentclass[12pt]{iopart}
\usepackage{iopams}

\usepackage{graphicx}
\usepackage{dcolumn}
\usepackage{bm}
\usepackage{comment}
\usepackage{multirow}
\usepackage{xcolor}
\usepackage{subfig}
\expandafter\let\csname equation*\endcsname\relax
\expandafter\let\csname endequation*\endcsname\relax
\usepackage{amsmath}

\begin{document}

\title{Analytical considerations for optimal axion haloscope design}

\author[kaist]{Junu Jeong$^1$, SungWoo Youn$^{1, *}$, Sungjae Bae$^{1,2}$, Dongok Kim$^{1, 2}$, Younggeun Kim$^{1, 2}$, Yannis K. Semertzidis$^{1,2}$}
\address{$^1$Center for Axion and Precision Physics Research, Institute for Basic Science, Daejeon 34051, Republic of Korea}
\address{$^2$Department of Physics, Korea Advanced Institute of Science and Technology, Daejeon 34141, Republic of Korea}
\ead{*swyoun@ibs.re.kr}
\vspace{10pt}
\begin{indented}
\item[]\today
\date{\today}
\end{indented}

\begin{abstract}
The cavity haloscope provides a highly sensitive method to search for dark matter axions in the microwave regime.
Experimental attempts to enhance the sensitivity have focused on improving major aspects, such as producing strong magnetic fields, increasing cavity quality factors, and achieving lowest possible noise temperatures.
Minor details, however, also need to be carefully considered in realistic experimental designs.
They are associated with non-uniform magnetic fields over the detection volume, noise propagation under attenuation and temperature gradients, and thermal disequilibrium in the cavity system.
We take analytical approaches to these topics and offer optimal treatments for improved performance.
\end{abstract}

\vspace{2pc}

\section{Introduction}
The axion is a hypothetical elementary particle resulting from the Peccei-Quinn mechanism proposed to solve the strong {\it CP} problem in particle physics~\cite{paper:PQ,paper:axion}.
The {\it invisible} axions with a small mass may have pervaded the early universe in a stable state~\cite{paper:invisible} and thus their relic abundance could provide a clue to the mysterious dark matter halo in our galaxy at the present time~\cite{paper:CDM}.
This pseudoscalar boson is regarded to be perceivable as a form of microwave photon to which it converts under a strong magnetic field~\cite{paper:haloscope}.
The cavity-based haloscope is the most sensitive and efficient experimental method to use when searching for dark matter axions as it detects the resonant signal power, which can be formulated as~\cite{paper:detection_rate}
\begin{equation}
    P_{a\gamma\gamma} = \frac{g_{a\gamma\gamma}^{2}\rho_{a}}{m_{a}^{2}}\omega_{c}\langle \mathbf{B}_{e}^{2}\rangle V G \frac{Q_{c}Q_{a}}{Q_{c} + Q_{a}},
    \label{eq:conv_power}
\end{equation}
where $g_{a\gamma\gamma}$ is the axion-photon coupling, $m_{a}$ and $\rho_{a}$ are the mass and local density of the dark matter axion, $\omega_{c}(=2\pi\nu)$ is the angular frequency of the cavity mode under consideration, $\langle \mathbf{B}_{e}^{2} \rangle$ is the average of the square of the externally applied magnetic field $\mathbf{B}_{e}$ inside the cavity volume $V$, and $Q_{c}$ and $Q_{a}$ are quality factors of the cavity and axion, respectively.
The form factor $G$ represents the geometric alignment of the cavity mode with the externally applied magnetic field and is given by
\begin{equation}
    G = \frac{|\int \mathbf{E}_{r}\cdot \mathbf{B}_{e} dV|^{2}}
    {\int \epsilon_r'|\mathbf{E}_{r}|^{2} dV \times \langle \mathbf{B}_{e}^{2}\rangle V},
    \label{eq:form_factor}
\end{equation}
where $\mathbf{E}_{r}$ is the axion-induced electric field associated with the resonant mode and $\epsilon_r=\epsilon_r' + i\epsilon_r''$ is the relative permittivity within the volume.

Since the mass of the axion is not predictable, the experiment must be capable of tuning the resonant frequency.
The performance of an experiment is characterized by the frequency scan speed with a given sensitivity, which is obtained with the radiometer equation~\cite{paper:dicke} using the Boltzmann constant $k_{B}$, as follows:
\begin{equation}
    \frac{d\nu}{dt} \simeq \left(\frac{1}{\textrm{SNR}} \right)^{2} \left(\frac{P_{a\gamma\gamma}}{k_{B}\mathbb{T}_{\textrm{sys}}}\right)^{2} \left(1 + \frac{Q_{a}}{Q_{c}}\right).
    \label{eq:scan_rate}
\end{equation}
Here, $\textrm{SNR}$ is the desired signal-to-noise ratio.
The system noise, represented by the equivalent temperature $\mathbb{T}_{\textrm{sys}}$, is determined from the linear sum of the thermal noise from the cavity and the added noise by the receiver chain, where the latter contribution is predominated by the noise of the first stage amplifier.
The search frequency is tuned by altering the electromagnetic field of the cavity mode, typically by employing a tuning structure inside the cavity.

Major experimental efforts to improve the scan speed have been made for the key elements~\cite{paper:axion_detect_review} -- high-field magnets~\cite{paper:25T} and large-volume high-quality cavities~\cite{paper:SCcavity} for signal enhancement, and low-temperature refrigerators and quantum-noise limited amplifiers~\cite{paper:axion_QNL,paper:HAYSTAC_Brubaker,paper:HAYSTAC_Zhong} for noise reduction.
However, minor ingredients also influence the signal and noise estimation, and thus have non-trivial effects on experimental performance.
They include non-uniformity of the applied magnetic field, noise development by lossy transmission lines with a temperature gradient, and thermal disequilibrium between the cavity and the tuning structure. 
These factors have been implicitly considered or partially addressed, but not explicitly discussed or fully addressed.
Herein, we review these topics in an analytical manner to examine the potential effects and provide optimal treatments for sensitivity improvement.

\section{Non-uniform magnetic fields}
\label{sec:cav}
The form factor, expressed here by Eq.~\ref{eq:form_factor}, is estimated based on a simulation and obtained in a straightforward manner under an uniform magnetic field.
In reality, however, the magnetic field of a solenoid with a finite length is not uniform over the magnet bore and gradually decreases along the main axis with the development of stray field in the radial direction.
Therefore, a reasonable estimation of $G$ requires knowledge of the overall field distribution within the cavity volume.

\begin{figure}
\centering
\includegraphics[width=0.6\linewidth]{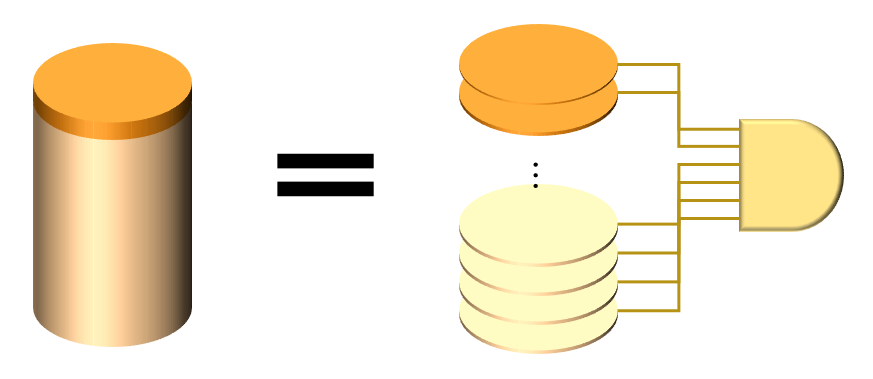}
\caption{A long cylindrical cavity is equivalent to a pileup of many disc-shaped cavities with individual signals combined by a power combiner.}
\label{fig:cavity_height_increase}
\end{figure}

As long as a magnetic field is present in space, the axion-to-photon conversions would take place and hence a longer cavity, which stores more of the magnetic field, would be beneficial.
However, Eqs.~\ref{eq:conv_power} and \ref{eq:form_factor} indicate that the conversion power is proportional to $\frac{|\int \mathbf{E}_{r}\cdot \mathbf{B}_{e} dV|^{2}}{\int |\mathbf{E}_{r}|^{2} dV}$, which eventually decreases with an increase in the cavity volume under a decreasing magnetic field.
This occurs because the energy density of axion-induced photons would not remain the same for a volume-varying cavity under a non-uniform magnetic field.
This effect can also be understood by approximating a single long cylindrical cavity under a non-uniform solenoid magnetic field to a series of disc-shaped cavities, each of which is under a uniform field with a different strength, coupled to a common signal combiner, such that increasing the cavity volume (length) is equivalent to adding more discs, as shown in Fig.~\ref{fig:cavity_height_increase}.
The individual disc cavities would generate different levels of signal power depending on the field strength.
Noting the fact that unbalanced inputs to a power combiner induce power dissipation, the combined power would be lower than a simple sum of the input powers~\cite{paper:combiner}.
In other words, a continuous increase of the cavity volume under a finite magnetic field would not always be beneficial.

This intuition indicates that the cavity length can be optimized for a given magnetic field distribution to maximize the scan speed, equivalently $\langle \mathbf{B}_{e}^{2} \rangle^{2}V^2G^2Q_{c}$, a product of the experimental parameters in Eq.~\ref{eq:scan_rate} with the limit of $Q_{c} \ll Q_{a}$.
For a cylindrical cavity, the quality factor is given by
\begin{equation}
    Q_{c} = \frac{1}{1 + R / L} \frac{R}{\delta},
    \label{eq:quality_factor}
\end{equation}
where $R$ and $L$ are correspondingly the cavity radius and length and $\delta$ is the skin depth of the cavity surface.
For the TM$_{010}$ mode, Eq.~\ref{eq:form_factor} can be expressed in the two-dimensional cylindrical coordinate system ($\rho, z$) as
\begin{equation}
    G = \frac{4}{\chi_{01}^{2}}\frac{B_{0}^{2}}{\langle \mathbf{B}_{e}^{2} \rangle}\left[\frac{\chi_{01}}{2J_{1}(\chi_{01})}\int J_{0}\left(\chi_{01}\frac{\rho}{R}\right)\mathbf{b}_e \cdot \hat{z} \frac{dV}{V} \right]^{2},
    \label{eq:form_factor_bessel}
\end{equation}
where $J_m$ is a $m$-th Bessel function of the first kind, $\chi_{01}$ is the first root of $J_0$, and $\mathbf{b}_e \equiv \mathbf{B}_{e} / B_{0}$ is the external magnetic field scaled by the field strength at the magnet center $B_0$.
For a uniform magnetic field, $\mathbf{b}_e = \hat{z}$, and the from factor reaches its maximum value of $4/\chi_{01}^{2} \approx 0.69$.
Eqs.~\ref{eq:quality_factor} and \ref{eq:form_factor_bessel} provide an expression of the scan speed in terms of the geometric parameters and functions, as
\begin{equation}
    \frac{d\nu}{dt} \propto V^{2} \left[\frac{RL}{R+L}\right]\left[ \int J_{0}\left(\chi_{01}\frac{\rho}{R}\right)\mathbf{b}_{e} \cdot \hat{z} \frac{dV}{V} \right]^{4}.
    \label{eq:scan_rate_geometric}
\end{equation}
The integral part in Eq.~\ref{eq:scan_rate_geometric} can be replaced by a series expansion of a known magnetic field profile at $\rho=0$, $b_{z}(z) \equiv \mathbf{b}_e(0, z)\cdot\hat{z}$, which yields an analytically resolvable form
\begin{equation}
    \frac{d\nu}{dt} \propto V^{2} \left[\frac{RL}{R+L}\right]\left[ \mathcal{C}_{1} \langle b_{z} \rangle - \frac{\mathcal{C}_{2}}{4} \langle R^{2}\partial_{z}^{2}b_{z} \rangle + \frac{\mathcal{C}_{3}}{64} \langle R^{4}\partial_{z}^{4}b_{z} \rangle + \cdots \right]^{4},
    \label{eq:scan_rate_expansion}
\end{equation}
where $\mathcal{C}_{i}$ is the coefficient of the series expansion (see Appendix~\ref{app:scan_speed} for the derivation).

\begin{figure}
\centering
\includegraphics[width=0.6\linewidth]{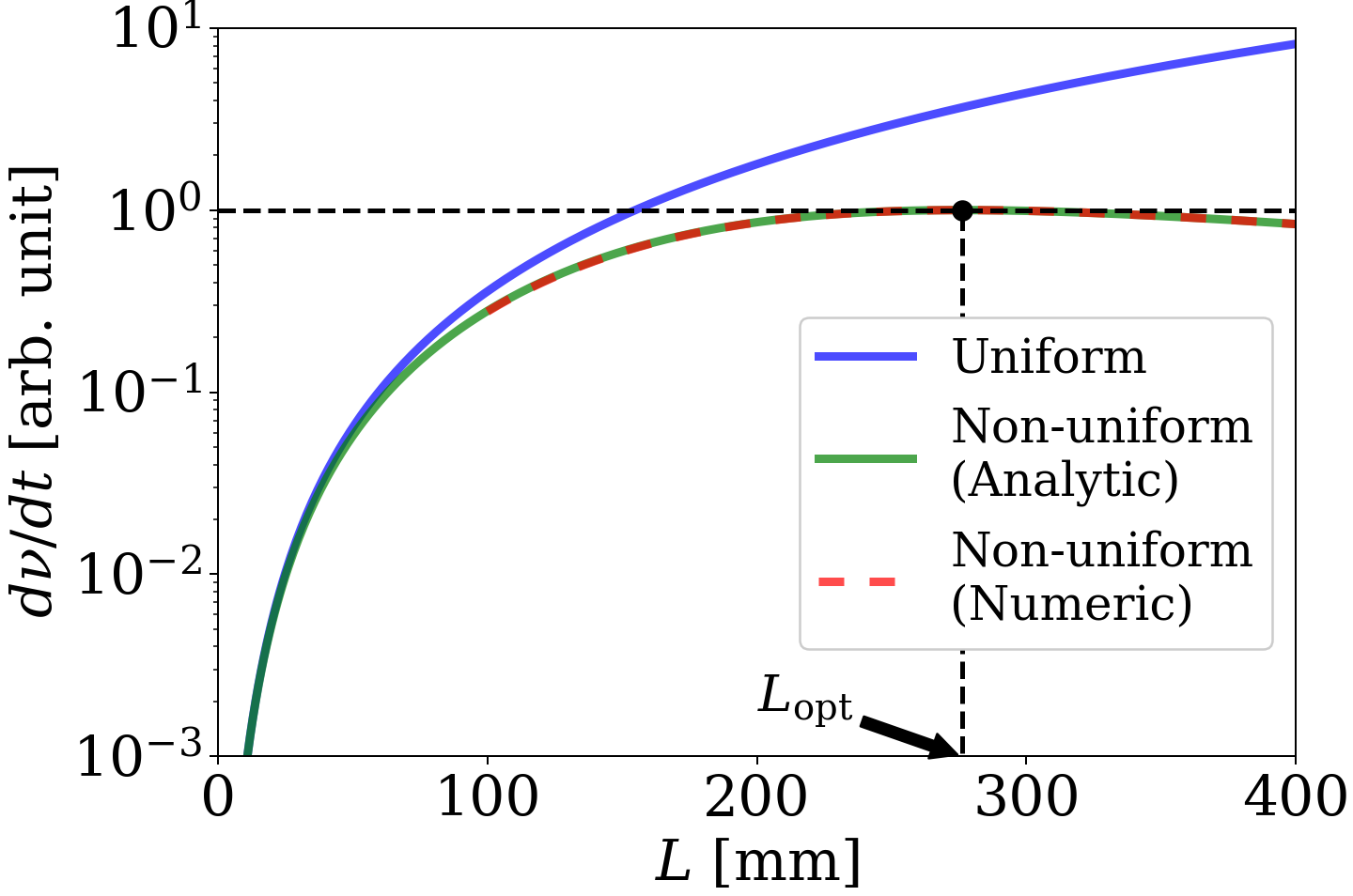}
\caption{\label{fig:dfdt_length} Scan speed calculated for a cavity experiment using a 9T/127mm SC magnet as a function of the cavity length. 
The quantity is normalized to its maximum value at the optimal length $L_{\rm opt}$.
The solid green and dashed red lines represent the analytical (Eq.~\ref{eq:scan_rate_expansion}) and numerical results, respectively, using the (non-uniform) magnetic profile provided by the manufacturer.
The blue line corresponds to the scan speed for a uniform magnetic field.}
\end{figure}

In practice, the magnetic field profile is available either from the manufacturer or by measurements, and the optimal cavity length can be obtained using Eq.~\ref{eq:scan_rate_expansion}.
An example is given by a cavity experiment reported in Ref.~\cite{paper:capp_9t}, where a 9T/127mm superconducting (SC) magnet was utilized.
Using the field profile provided by the manufacturer, Eq.~\ref{eq:scan_rate_expansion} is solved as a function of the cavity length, as shown in Fig.~\ref{fig:dfdt_length}.
The analytical calculation is in very good agreement with a simulation result based on the finite element method.
The optimal cavity length $L_{\rm opt}$ is found at the maximum value of the scan speed by requiring $\partial_{L}\left(d\nu/dt\right)=0$.
It is noted that uniform magnetic fields yield considerable high scan speeds particularly for large cavity lengths, which could lead to an overestimation of the experimental sensitivity.

\begin{figure}
\centering
\includegraphics[width=0.6\linewidth]{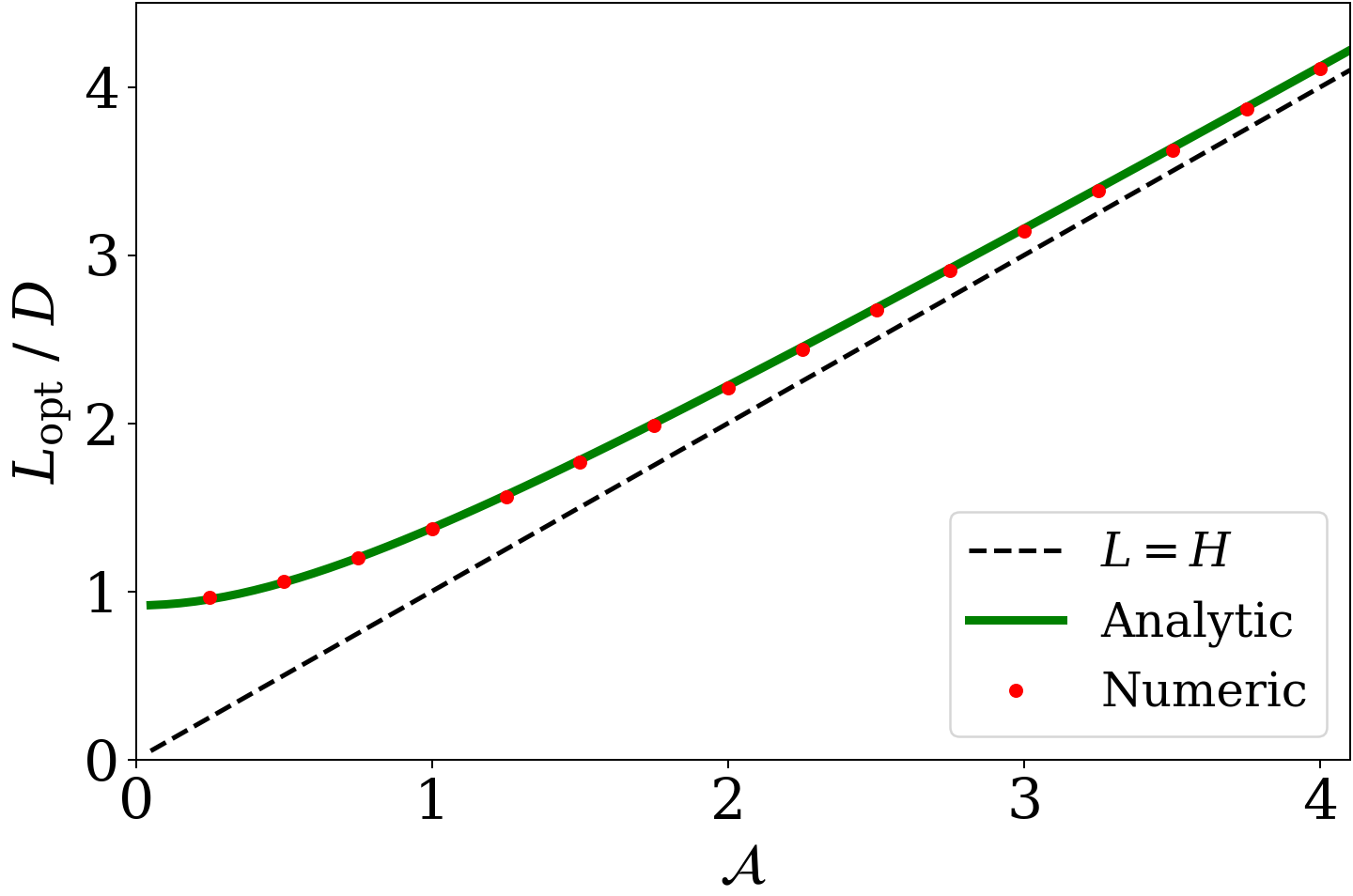}
\caption{\label{fig:optimal_length}Optimal length of a cylindrical cavity obtained using $\partial_{L}\left(d\nu/dt\right)=0$ from Eq.~\ref{eq:scan_rate_expansion} as a function of the aspect ratio of a solenoid.
Numerical calculations for various values of the aspect ratio are overlapped.}
\end{figure}

The optimal length of a cylindrical cavity, in general, depends on the relative dimension of the given magnet, typically represented by the aspect ratio $\mathcal{A}=H/D$, where $H$ and $D$ are correspondingly the magnet height and the bore diameter.
The dependency obtained based on Eq.~\ref{eq:scan_rate_expansion} is presented in Fig.~\ref{fig:optimal_length}, where the overall behavior is verified by numerical calculations for various values of $\mathcal{A}$.
Some interesting features are noticed: 1) the optimal cavity length is always larger than the magnet height; and 2) it converges to the magnet height for large values of $\mathcal{A}$ while approaching the bore size for small values.
This provides a guideline for the cavity design for a given magnet.

\section{Noise propagation}\label{sec:noise}

One of the intricate tasks in axion experiments is to estimate the system noise, particularly the added noise caused by the receiver chain, with high accuracy.
This is true because a typical cryogenic cooling system consists of multiple stages with RF components mounted, making it complicated to understand how the noise propagates between components at different temperatures.
The RF signal and noise go through attenuation as well as a temperature change while propagating through the components.
These effects influence the accuracy of the relevant measurements and thus need to be properly understood.

A non-zero temperature $T$ generates thermal noise power of $P_n = k_B T \Delta\nu$ over bandwidth $\Delta\nu$~\cite{paper:thermal_noise}.
When thermal noise propagates between two different physical temperatures, from $T_i$ to $T_f$, through a passive component with attenuation $A$, the noise power at the final stage is given by the Johnson-Nyquist formula as
\begin{equation}
\begin{aligned}
    P_n &= k_B T_i A \Delta \nu + k_B T_f \Delta \nu (1-A)\\
    &= k_B \Delta \nu \left[T_i A + T_f (1-A) \right].
\end{aligned}
\label{eq:noise_prop}
\end{equation}
Eq.~\ref{eq:noise_prop} reads that the resistor generates the same amount of noise as that attenuated by itself according to the law of conservation of energy.
The variable in the square brackets is defined as the equivalent noise temperature,
\begin{equation}
    \mathbb{T} \equiv T_i A + T_f (1-A),
    \label{eq:noise_temp}
\end{equation}
which satisfies $\mathbb{T} = T$ when $T_i = T_f=T$.

\begin{figure}
\centering
\includegraphics[width=0.6\linewidth]{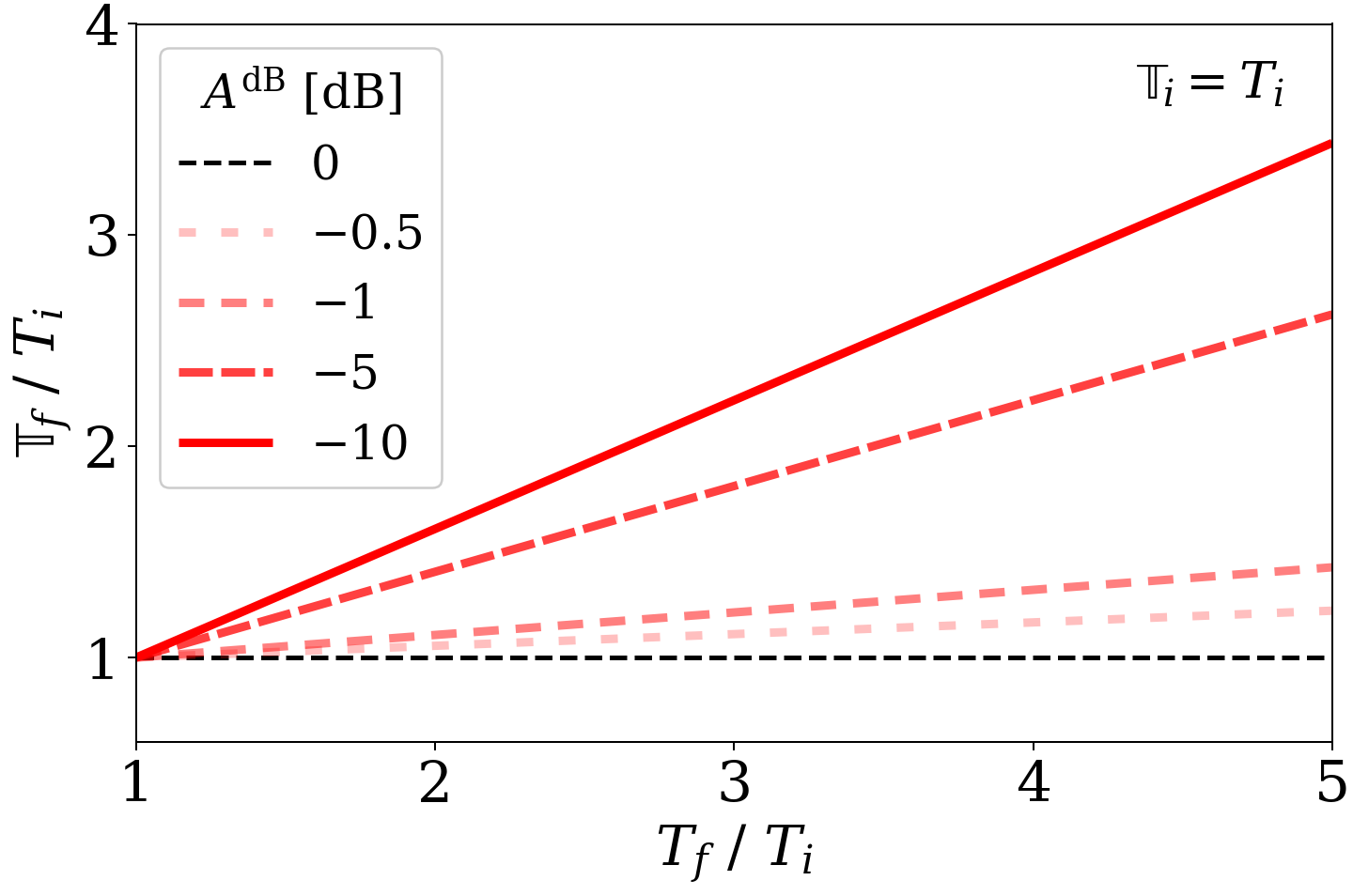}
\caption{\label{fig:noise_prop}Development of noise temperature of propagating noise obtained from Eq.~\ref{eq:noise_temp_exp} with increasing the physical temperature at the final stage (normalized to the initial temperature) for different values of attenuation. 
Here, $A^{\rm dB}$ specifies the attenuated values in decibels.}
\end{figure}

For general components, RF propagation is associated with {\it gradual} attenuation and temperature variations.
A good example is a lossy transmission line connecting two different stages of a cooling system.
Such a component can be approximated as a continuum of $N$ identical small segments, each of which has constant attenuation $\delta A$ and constant temperature variation $\delta T$, such that $A=(\delta A)^N$ and $T_f - T_i = N\delta T$.
The noise temperature at the $n$-th segment is then given by
\begin{equation*}
    \mathbb{T}_{n} = \mathbb{T}_{n-1} \delta A + T_{n}(1 - \delta A).
\end{equation*}
At the limit of $N\rightarrow\infty$, Eq.~\ref{eq:noise_temp} can be reformulated to
\begin{equation}
\label{eq:noise_temp_exp}
    \mathbb{T}_{f} = \mathbb{T}_{i} + (T_{f} - T_{i})\left(1 + \frac{1 - A}{\ln A} \right) + (T_{i} - \mathbb{T}_{i})(1 - A),
\end{equation}
which describes noise propagation in a more realistic sense (see Appendix~\ref{app:noise_prop} for the derivation).
Figure~\ref{fig:noise_prop} shows the dependence of the propagated noise temperature on the difference between the initial and final physical temperatures for various values of attenuation.
The effective noise is elevated by a positive temperature gradient while being depressed by attenuation.

The effects of the temperature gradient and attenuation on the noise propagation must be carefully taken into account for an accurate estimation of the system noise.
For cavity haloscope experiments, there are two circumstances where such effects would matter.
First, noise measurements typically rely on the responses of the device under test (DUT) to its inputs.
For instance, for the Y-factor method, a noise source feeds the thermal noise equivalent to two different temperatures through a non-lossless RF line into the DUT, whose temperature is held constant during the measurements.
Second, the thermal noise generated from a detection cavity undergoes attenuation by a transmission line before the first stage of the receiver chain, which can be placed at a different temperature.
We use Eq.~\ref{eq:noise_temp_exp} to treat the noise propagation appropriately and to estimate the noise temperatures in realistic conditions.

\subsection{Y-factor method}

The Y-factor method, a popular technique for an accurate measurement of the noise figure of a DUT, is widely adopted in axion search experiments to estimate the gain and noise temperature of an amplifier or an entire receiver chain~\cite{note:Y-factor}.
This method evaluates the intrinsic noise of the DUT based on the ratio of the output powers to the input thermal noise powers corresponding to two different known temperatures.
The noise source could be a diode which has a pre-calibrated excess noise ratio (ENR) or a resistor whose temperature can be controlled by a heater~\cite{paper:noise_measure}.

If the output noise powers through the DUT for two different (cold and hot) source temperatures, $T_c$ and $T_h$, are denoted by $P_c$ and $P_h$, respectively, the noise temperature of the DUT itself can be obtained using the following well-known equations:
\begin{equation}
    \mathbb{T}_{\rm DUT} = \frac{T_h - Y T_c}{Y - 1}, \quad Y = \frac{P_h}{P_c}.
    \label{eq:y-factor}
\end{equation}
However, in reality, the thermal noise from the source undergoes attenuation and a temperature gradient while propagating through a transmission line to the input of the DUT.
This effect can be reflected by replacing $T_{h,c}$ with $\mathbb{T}_{h,c}$ in Eq.~\ref{eq:y-factor}; i.e.,
\begin{equation}
    \mathbb{T}_{\rm DUT} = \frac{\mathbb{T}_{h} - Y \mathbb{T}_{c}}{Y - 1}, 
    \label{eq:y-factor_exp}
\end{equation}
where $\mathbb{T}_{h,c}$ obeys Eq.~\ref{eq:noise_temp_exp}.
For a noise source initially in a thermal equilibrium with the DUT, $\mathbb{T}_c=T_c$ holds, as can be verified by Eq.~\ref{eq:noise_temp_exp}.

\begin{figure}
\centering
\includegraphics[width=0.6\linewidth]{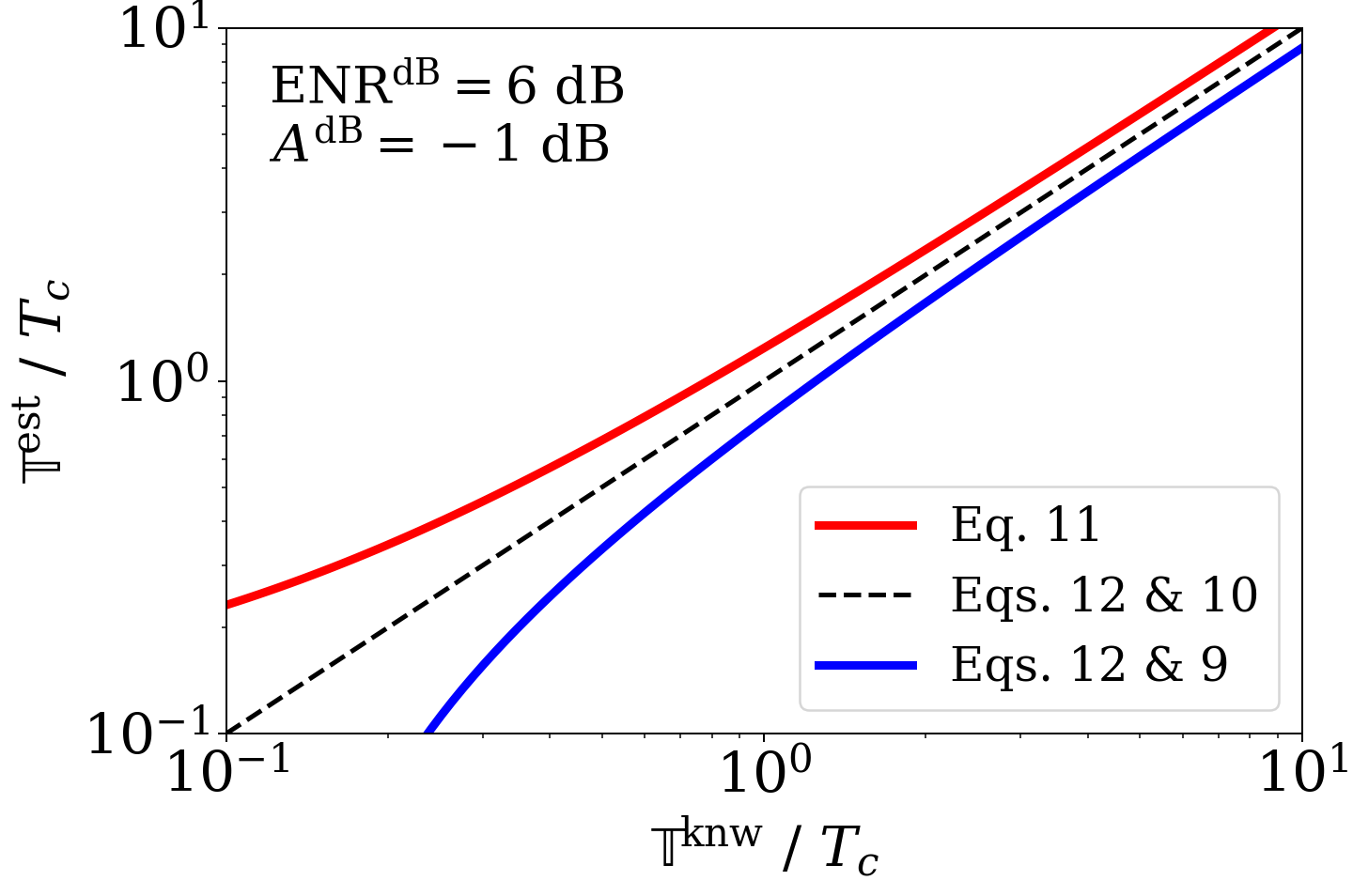}
\caption{\label{fig:yfactor_error}Estimated noise temperatures ($\mathbb{T}^{\rm est}$) of DUTs with known noise figures ($\mathbb{T}^{\rm knw}$) using the Y-factor method for different treatments of the noise propagation. 
$\mathbb{T}^{\rm knw}$ is normalized to $T_c$, and ${\rm ENR^{dB}}=6$\,dB and line attenuation of $A_{\rm lin}^{\rm dB}=-1$\,dB are assumed here.
The black line represents Eq.~\ref{eq:y-factor_exp} with Eq.~\ref{eq:noise_temp_exp} where the noise propagation is appropriately taken into account.
The effects of an improper treatment of noise estimation are visualized by the red and blue lines, which are obtained from Eq.~\ref{eq:y-factor}, where the physical temperatures are used, and Eq.\ref{eq:y-factor_exp} with Eq.~\ref{eq:noise_temp}, where the attenuation is described as a single-step function, respectively.
}
\end{figure}

Figure~\ref{fig:yfactor_error} exhibits the effect of the noise propagation on the noise estimation using the Y-factor method.
Without consideration of the effect, equivalent to Eq.~\ref{eq:y-factor}, the Y-factor method overestimates the noise of the DUT, as represented by the red line.
The effect would be manifested when a DUT is placed at a cryogenic temperature while a noise diode is placed at room temperature as such a configuration would typically add more attenuation and a higher temperature gradient.
On the other hand, if the attenuation is not properly considered, e.g., if it is described by a single-step function as Eq.~\ref{eq:noise_temp}, the Y-factor method yields underestimated noise figures, as represented by the blue line.
It is also noted that the discrepancies are greater for DUTs with lower noise levels.
Therefore, the noise propagation must be carefully considered when estimating the noise properties of the receiver chain in a complicated system.

\subsection{Cavity thermal noise}

In axion haloscopes, the total noise of the system is evaluated as the thermal noise generated by the cavity transmitted through a RF line to the receiver chain, where the shot noise of electronics is linearly added while propagating through the chain.
From the argument developed in this section, the noise propagation effects must be reflected when estimating the noise temperature of the system such that
\begin{equation}
    \mathbb{T}_{\rm sys} = \mathbb{T}_{\rm thr} + \mathbb{T}_{\rm add},
    \label{eq:system_noise_ref}
\end{equation}
where $\mathbb{T}_{\rm thr}$ is the equivalent temperature of the thermal noise traveling from the cavity to the first-stage amplifier, obtained using Eq.~\ref{eq:noise_temp_exp}, while $\mathbb{T}_{\rm add}$ is the added noise temperature stemming from the receiver chain, obtained from Eq.~\ref{eq:y-factor_exp}.

\begin{figure}
\centering
\subfloat[\label{fig:noise_prop_config}]{\includegraphics[width=0.45\linewidth]{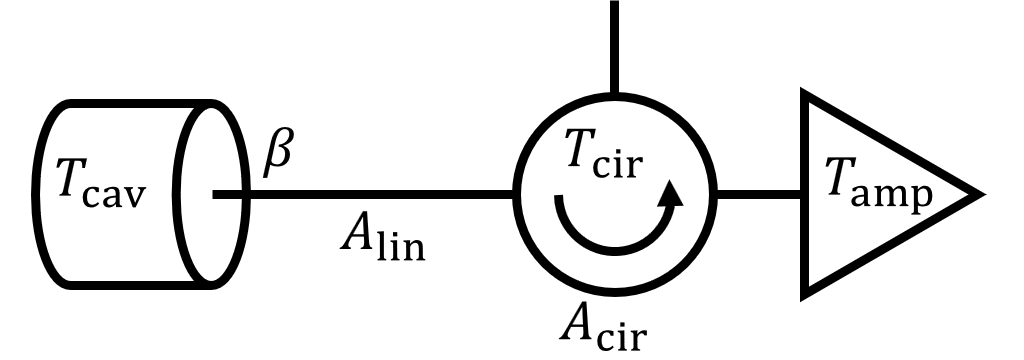}}
\vspace{0.2cm}
\subfloat[\label{fig:noise_prop_Tamp}]{\includegraphics[width=0.6\linewidth]{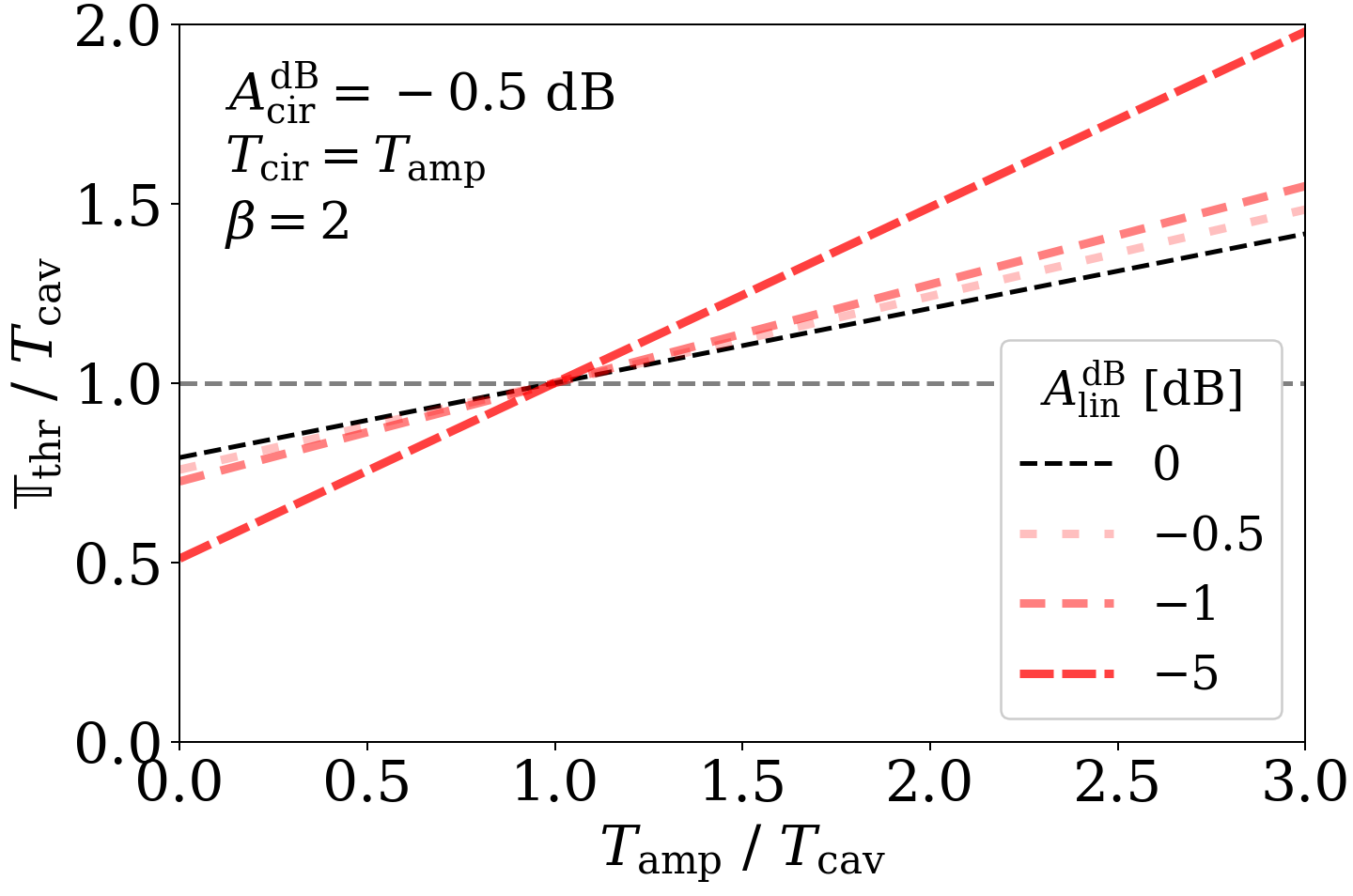}}
\caption{(a) Schematic diagram of a cavity haloscope composed of (left to right) a cavity, a circulator, and an amplifier.
(b) Equivalent noise temperature seen by the amplifier as a function of its physical temperature relative to that of the cavity for different values of the line attenuation $A_{\rm lin}^{\rm dB}$.
The effects of the addition of the circulator ($A_{\rm cir}^{\rm dB}=-0.5$\,dB) and an over-coupled antenna ($\beta=2$) are reflected in Eq.~\ref{eq:noise_temp_seen_amp}.
The absence of the circulator and line attenuation is represented by the grey dotted line.}
\end{figure}

In a more practical haloscope design, additional RF components could be introduced between the detection cavity and the receiver chain for multiple purposes.
A typical example is a RF circulator, which serves to circumvent the impedance mismatching effect and/or to measure the cavity properties.
As illustrated in Fig.~\ref{fig:noise_prop_config}, it could be placed at a different temperature.
To maximize the scan speed, the signal pickup antenna is configured to over-couple to the cavity~\cite{paper:detection_rate}, conventionally represented by $\beta>1$, where $\beta$ is the coupling strength.
In such a configuration, not only the signal but also the noise is subjected to reflection due to the impedance mismatch between the antenna and the cavity.
This complicates the noise propagation, which is described as 
\begin{equation}
\label{eq:noise_temp_seen_amp}
    \mathbb{T}_{\rm thr} = \frac{4 \beta}{\left( 1 + \beta \right)^{2}}\mathbb{T'}_{\rm thr} + \left(\frac{1 - \beta}{1 + \beta}\right)^{2}\mathbb{T''}_{\rm thr},
\end{equation}
where $'$ represents direct propagation of the noise originating from the cavity to the receiver, while $''$ represents the back-and-forth propagation of the noise coming from the circulator, reflected by the cavity, and going to the receiver.
Eq.~\ref{eq:noise_temp_seen_amp} can be used to estimate the noise temperature from the perspective of the receiver (the first-stage amplifier), depending on the physical temperature of the amplifier relative to that of the cavity.
Figure~\ref{fig:noise_prop_Tamp} gives an example of such a dependency in the presence of a circulator between the two components for various values of the line attenuation.
It would be beneficial to maintain the physical temperature of the amplifier lower than that of the cavity.

\section{Hot rod problem}\label{sec:hotrod}

It was reported in an axion haloscope experiment that the noise power near the cavity resonant frequency was significantly higher than the noise power observed off resonance~\cite{paper:HAYSTAC_Brubaker}.
It turns out that the excess noise on resonance was attributed to poor thermalization of the frequency tuner (a conducting rod) located inside the cavity.
Known as the hot rod problem, the issue was addressed to some extent by improving the thermal link to the tuner at the cost of a reduction of the cavity quality factor~\cite{paper:HAYSTAC_Zhong}.
As such an issue could be common in haloscope experiments, it should be carefully considered when designing an experiment.
Here we analytically estimate the effect of the thermal disequilibrium between the cavity and the frequency tuner on the overall thermal noise.

Many axion experiments are conducted at very low temperatures due to the advanced cryogenic technologies, necessitating consideration of the quantum effects in thermal noise.
In this regard, we define the effective temperature as follows:
\begin{equation}
\mathcal{T}_{\rm eff} \equiv \frac{\hbar\omega} {k_{B}} \left( \frac{1}{e^{\hbar\omega/ k_{B}T_{\rm phy}} - 1}+\frac{1}{2} \right).
\label{eq:eff_temp}
\end{equation}
Here, $\hbar$ is the reduced Planck constant.
The first term of Eq.~\ref{eq:eff_temp} represents the average thermal photon number at physical temperature $T_{\rm phy}$ at frequency $\omega$, while the second term accounts for zero-point fluctuations, reflecting the quantum limit near absolute zero.

For a cavity with a tuning structure inserted, the thermal temperature observed by the antenna coupled to the cavity can be decomposed into two components:
\begin{equation*}
\label{eq:noise_portion}
    \mathcal{T}_{\rm obs} = f_{\rm cav} \mathcal{T}_{\rm cav} + f_{\rm tun} \mathcal{T}_{\rm tun}, \quad f_{\rm cav} + f_{\rm tun} = 1,
\end{equation*}
where $\mathcal{T}_{\rm cav,tun}$ are the effective temperatures of the cavity and the tuning structure, with $f_{\rm cav, tun}$ representing their fractional contributions.
Given that, as previously mentioned, passive devices generate noise to the extent of their dissipation, the fractions can be written as
\begin{equation}
\label{eq:noise_portion}
    f_{\rm cav} = \frac{P_{\rm cav}}{P_{\rm tot}} \quad {\rm and} \quad f_{\rm tun} = \frac{P_{\rm tun}}{P_{\rm tot}},
\end{equation}
where $P_{\rm cav, tun}$, satisfying $P_{\rm tot} = P_{\rm cav} + P_{\rm tun}$, are the dissipated power by the corresponding objects.
Depending on the material, the power dissipation formulae for an arbitrary resonant mode with electromagnetic fields of $\mathbf{E}$ and $\mathbf{H}$ are given by
\begin{equation}
\begin{split}
    & P_{de} = \frac{\omega_{c}\epsilon_0}{2} \int_{V}\epsilon_{r}'' |\mathbf{E}|^{2} dV, \\
    & P_{cd} = \frac{R_s}{2} \int_{S}|\mathbf{H}|^{2} dS,
\end{split}
\label{eq:power_diss}
\end{equation}
where the subscripts $de$ and $cd$ denote the dielectrics and conductors.
For the former, the power dissipation is associated with the dielectric loss $\epsilon_{r}''$ within its volume, while for the latter, it depends on the surface resistance $R_s=\frac{1}{\sigma\delta}$ (with conductivity $\sigma$) over its surface area.

\begin{figure}
\centering
\subfloat[\label{fig:noise_portion_1}]{\includegraphics[width=0.6\linewidth]{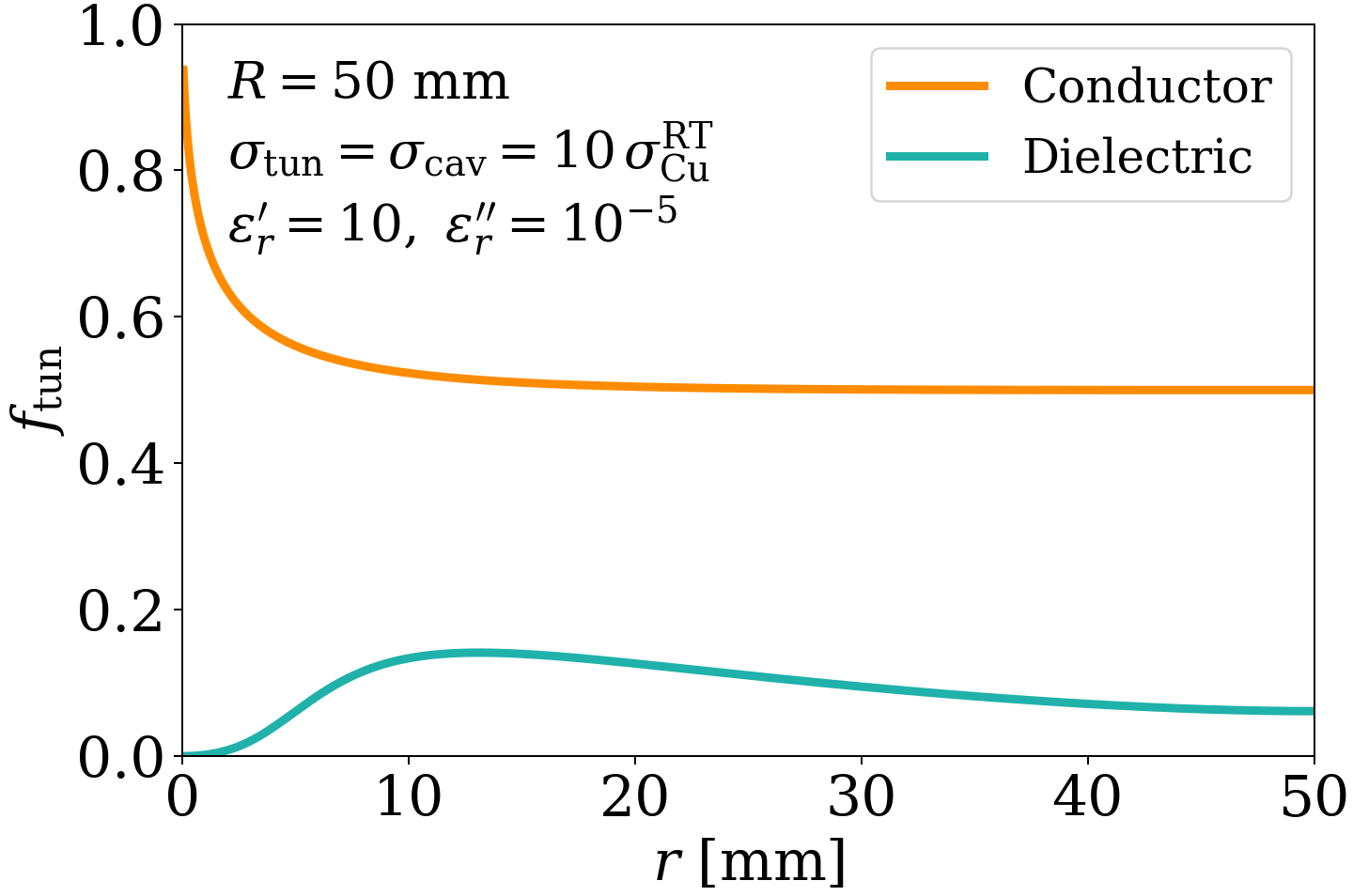}}
\vspace{0.1cm}
\subfloat[\label{fig:noise_portion_2}]{\includegraphics[width=0.6\linewidth]{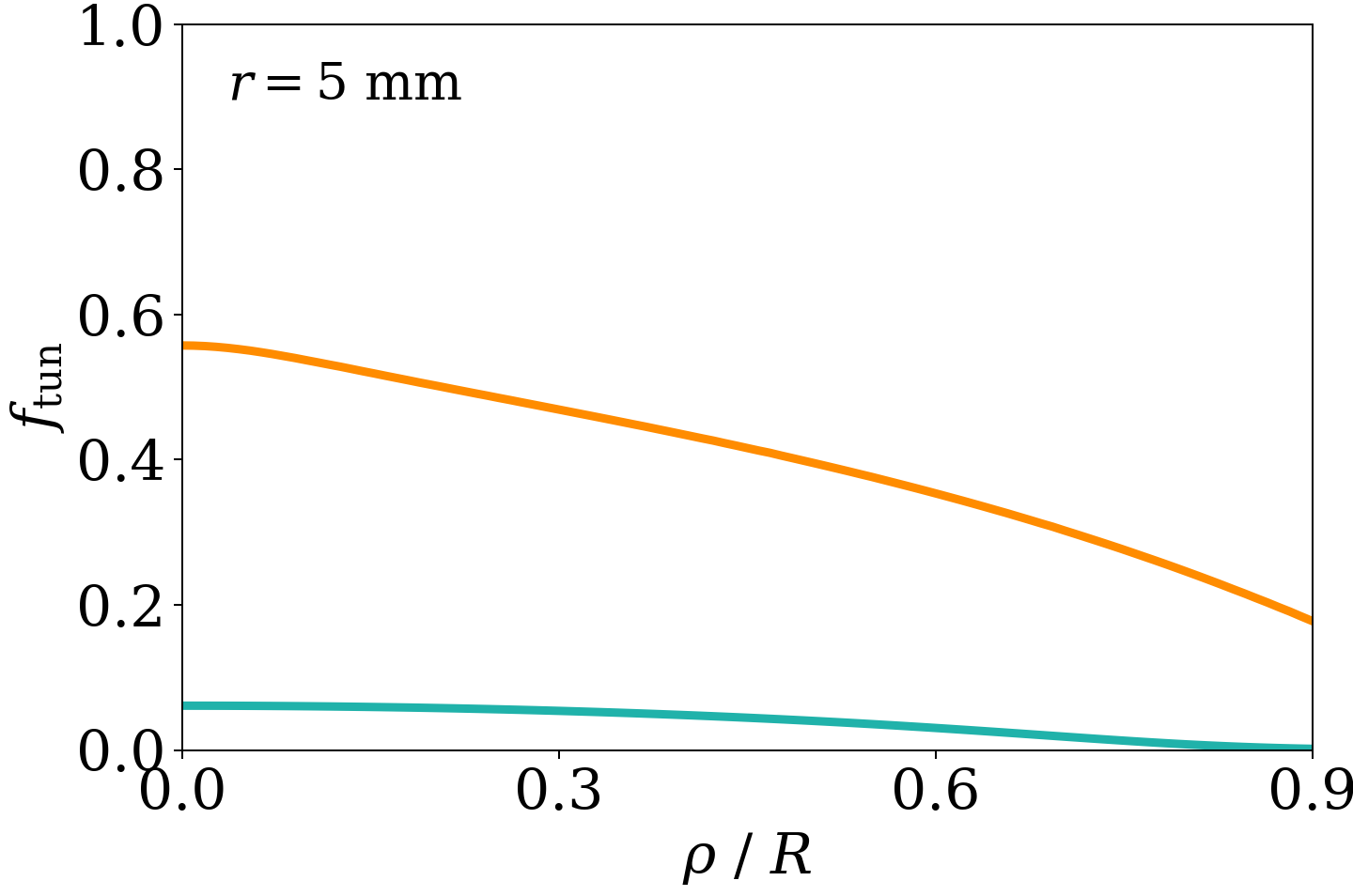}}
\caption{Fractional contributions by a cylindrical tuner inside a cylindrical cavity depending on (a) the size and (b) the position relative to the cavity center, $\rho$, for the given dimensions and properties written on the plots.
Two different materials, the conductor and dielectric, are represented by the orange and green lines, respectively.}
\label{fig:noise_portion}
\end{figure}

\begin{figure}
\centering
\includegraphics[width=0.6\linewidth]{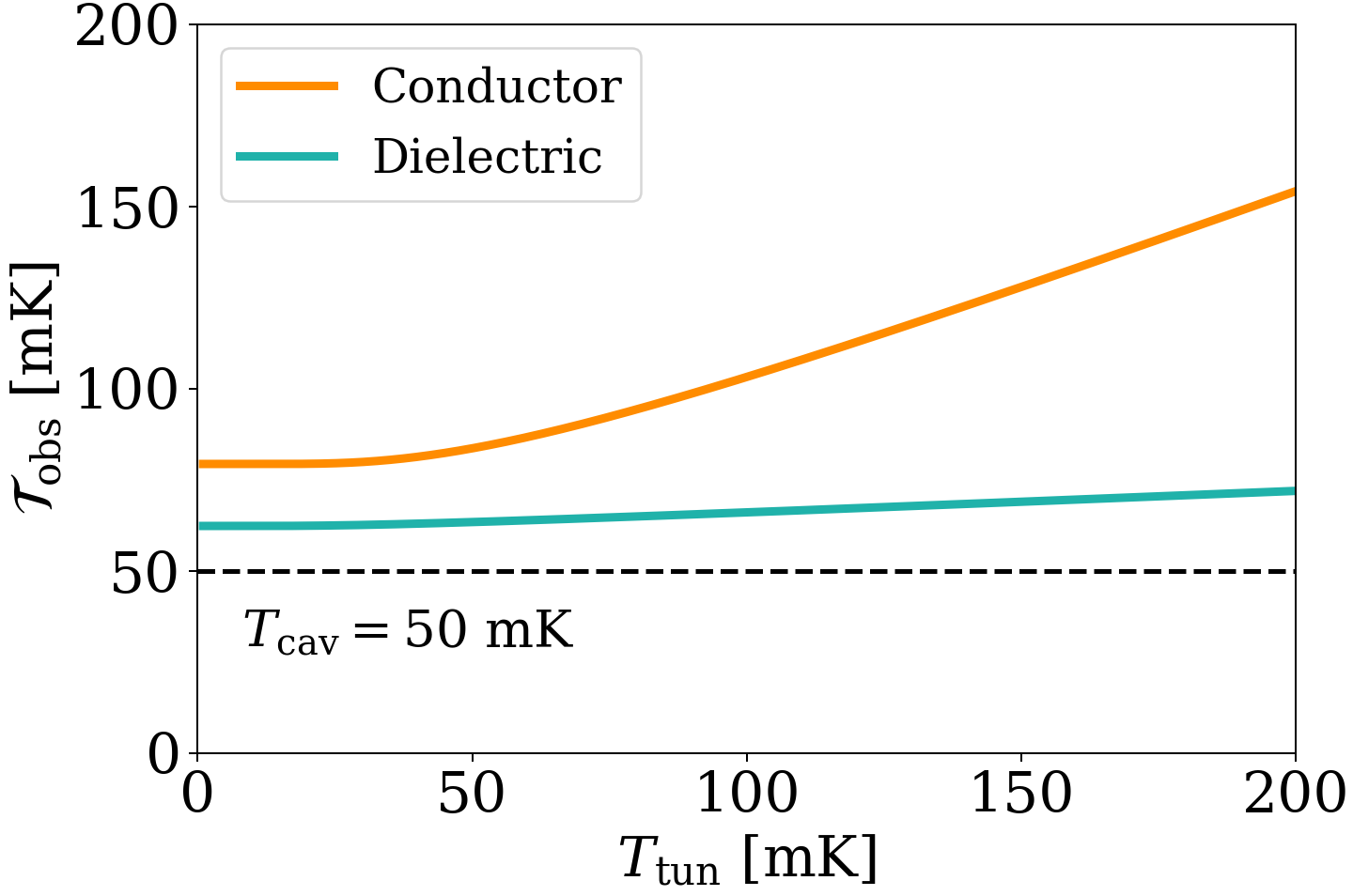}
\caption{\label{fig:noise_temp_for_rod_temp}Observed effective temperatures from a cavity placed at 50\,mK as a function of the physical temperature of a tuning rod located at the cavity center. The geometric and physical properties of the system are identical to those in Fig.~\ref{fig:noise_portion}. The quantum effects are taken into account (Eq.~\ref{eq:eff_temp}).}
\end{figure}

Equation~\ref{eq:power_diss} can be explicitly solved for a certain configuration, for instance, a long cylindrical cavity of radius $R$ with a cylindrical conducting rod of radius $r$ positioned at the center of the cavity.
The dissipated power by the cavity and rod is proportional to $2\pi R |\mathbf{H}(R)|^2$ and $2\pi r |\mathbf{H}(r)|^2$, respectively, and Eq.~\ref{eq:noise_portion} simply becomes
\begin{equation}
    f_{\rm cav,tun} = \frac{R|\mathbf{H}(R)|^2, r|\mathbf{H}(r)|^2}{R|\mathbf{H}(R)|^2 + r|\mathbf{H}(r)|^2},
\end{equation}
which is verified using the field profiles obtained from simulations with various rod radii.
The fractional contributions from the tuning rods of two different materials are shown in Fig.~\ref{fig:noise_portion_1} as a function of the relative rod size $r/R$.
Figure~\ref{fig:noise_portion_2} exhibits the dependence of the noise contribution on the rod position with respect to the center of the cavity; the closer to the cavity center it is, the larger the contribution it makes.
Figure~\ref{fig:noise_temp_for_rod_temp} provides an example showing how the observed effective temperature would develop with different rod temperatures assuming a configuration identical to that in Fig.~\ref{fig:noise_portion_1} with $r/R=0.1$.
Because, depending on detector design, the noise contribution by a tuner can be substantial, especially for a conductor, it would be essential to establish a good thermal link to it.

\section{Conclusion}

In this report, we considered some minor issues to be addressed in axion haloscope experiments.
We explicitly analyzed their influences on signal and noise estimation, based on which we suggest optimal designs to improve search performance.
Non-uniformities in external magnetic fields limits the energy density of axion-induced photons for a given cavity, and an optimal length was found to maximize the signal power.
The noise propagation was practically evaluated under temperature gradient and RF attenuation conditions to derive appropriate treatments for noise estimations using the Y-factor method and noise development with additional components.
In addition, the hot rod problem was revisited to provide an analytical method by which to calculate the effects of thermal disequilibrium between a cavity and a frequency tuner.
These effects need to be carefully taken into account when designing or conducting an experiment, as they will have non-trivial effects on experimental sensitivity.

\section*{Acknowledgement}
This work was supported by the Institute for Basic Science (IBS-R017-D1-2021-a00).

\appendix

\section{Series expansion of the form factor}
\label{app:scan_speed}
The electric field solution for the TM$_{010}$ mode of a cylindrical cavity with radius $R$ is given in the cylindrical coordinates $(\rho,\phi,z)$ with time $t$ by
\begin{equation*}
    \mathbf{E}_{\textrm{TM}_{010}} = \mathcal{E} J_{0}\left(\chi_{01}\frac{\rho}{R} \right) e^{-i \omega_{c}t}\hat{z},
\end{equation*}
where $\mathcal{E}$ is the mode amplitude, $\omega_{c}$ is the resonant frequency of the mode, $J_0$ is the zeroth Bessel function of the first kind and $\chi_{01}$ is its first root.
Using this solution, Eq.~\ref{eq:form_factor} can be transformed to 
\begin{equation}
    G = \frac{4}{\chi_{01}^{2}}\frac{B_{0}^{2}}{\langle \mathbf{B}_{e}^{2} \rangle}\left[\frac{\chi_{01}}{2J_{1}(\chi_{01})}\int J_{0}\left(\chi_{01}\frac{\rho}{R}\right)\mathbf{b}_{e} \cdot \hat{z} \frac{dV}{V} \right]^{2},
    \label{eq:form_factor_appendix}
\end{equation}
where $\mathbf{b}_{e} \equiv \mathbf{B}_{e} / B_{0}$ is the scaled magnetic field, which becomes $\mathbf{b}_{e} = \hat{z}$ for an uniform magnetic field.
Equation~\ref{eq:form_factor_appendix} expresses the form factor as a distribution function of $\mathbf{b}_{e}$.
Employing the Ampere's law in vacuum and the divergence-free condition of the applied magnetic field, we can expand $\mathbf{b}_e\cdot\hat{z}$ in power series of the field profile along the solenoid axis, $b_{z}(z) \equiv \mathbf{b}_{e}(\rho=0, z)\cdot\hat{z}$, as
\begin{equation*}
    \begin{split}
    \mathbf{b}_{e}(\rho, z)\cdot \hat{z} =& J_{0}(\rho\partial_{z}) b_{z}(z),\\
    =& b_{z}(z)-\frac{1}{4}\rho^{2}\partial_{z}^{2}b_{z}(z) + \frac{1}{64}\rho^{4}\partial_{z}^{4}b_{z}(z) - \cdots,
\end{split}
\end{equation*}
The integral part in Eq.~\ref{eq:form_factor_appendix} becomes
\begin{equation*}
    \mathcal{C}_{1}\langle b_{z} \rangle - \frac{\mathcal{C}_{2}}{4} \langle R^{2}\partial_{z}^{2}b_{z} \rangle + \frac{\mathcal{C}_{3}}{64} \langle R^{4}\partial_{z}^{4}b_{z} \rangle - \cdots,
\end{equation*}
where $\langle \rangle$ represents the average value over the cavity length for the relevant quantity.
The coefficient $\mathcal{C}_{i}$ is obtained from a volume integral of the Bessel function in Eq.~\ref{eq:form_factor_appendix} as follows
\begin{equation*}
    \begin{split}
        \mathcal{C}_{1} =&1, \\   
        \mathcal{C}_{2} =& \frac{2J_{2}(\chi_{01})-J_{3}(\chi_{01})\chi_{01}}{J_{1}(\chi_{01})\chi_{01}} \approx 0.30834, \\
        \mathcal{C}_{3} =& \frac{J_{3}(\chi_{01})(8-\chi_{01}^{2})}{J_{1}(\chi_{01})\chi_{01}^{2}} \approx 0.146935, \\
        & \vdots \\
        \mathcal{C}_{n} =& \frac{\chi_{01}}{J_{1}(\chi_{01})} \frac{{}_{1}F_{2} (n; 1, 1+n; -\chi_{01}^{2}/4)}{2n}
        \sim \frac{1.55}{\left(n - 1 + \sqrt{1.55}\right)^{2}},
    \end{split}
\end{equation*}
where ${}_{1}F_{2}(a_{1};b_{1},b_{2};x)$ is a generalized hypergeometric function~\cite{book:hypergeometric}.

\section{General equation for noise propagation}\label{app:noise_prop}

In axion haloscopes, signal and noise go through power loss and temperature difference through a series of RF components at different stages.
To a first approximation, a typical RF component has gradually varying attenuation and temperature, and thus they can be depicted as a continuum of $N$ identical small segments, $n$-th of which has infinitesimal attenuation $\delta A_n$ and temperature variation $\delta T_n$, such that 
\begin{equation}
\begin{split}
    A_n/A_{n-1} = \delta A_n, &\quad T_n-T_{n-1} = \delta T_n, \\
    A_n = \prod_{m=1}^n \delta A_m, \ A_N=A, &\quad T_n -T_0 = \sum_{m=1}^n \delta T_m.
    \label{eq:noise_prop_condition}
\end{split}
\end{equation}
Applying Eq.~\ref{eq:noise_temp}, the noise temperature at the $n$-th segment, $\mathbb{T}_{n}$, is given by
\begin{equation}
\label{eq:noise_prop_n}
    \mathbb{T}_{n} = \mathbb{T}_{n-1} \delta A_n + T_{n}(1 - \delta A_n),
\end{equation}
which must be distinct from the physical temperature $T_n$.
If we assume constant attenuation and temperature gradient for all segments, i.e., $\delta A_n=\delta A$ and $\delta T_n = \delta T$ for all $n$, then Eq.~\ref{eq:noise_prop_n} becomes
\begin{equation}
    \mathbb{T}_{n} = \mathbb{T}_{n-1}\delta A + (T_0 + n \delta T)(1 - \delta A).
\label{eq:noise_prop_recursive}
\end{equation}
This assumption would reasonably hold particularly for commercial RF transmission cables connecting two different temperature stages. 

Solving the recurrence relation Eq.~\ref{eq:noise_prop_recursive} for infinite $N$ with the initial condition of $\mathbb{T}_{0} = T_{0}$, we obtain
\begin{equation*}
    \mathbb{T}_{n} = T_{0} + n \delta T - \delta A \delta T \frac{1 - (\delta A)^{n}}{1 - \delta A},
\end{equation*}
which eventually yields an expression for the total noise temperature 
\begin{equation}
    \mathbb{T}_{N} = T_{N} + (T_{N} - T_{0}) \frac{1 - A}{\ln A},
\end{equation}
in terms of the known total attenuation $A$ and temperature variation $T_N-T_0$.
For the initial condition of $\mathbb{T}_{0} \neq T_{0}$, the solution for Eq.~\ref{eq:noise_prop_recursive} returns Eq.~\ref{eq:noise_temp_exp}
\begin{equation}
\label{eq:noise_prop_appendix}
    \mathbb{T}_{N} = \mathbb{T}_{0} + (T_{N} - T_{0})\left(1 + \frac{1 - A}{\ln A} \right) + (T_{0} - \mathbb{T}_{0})(1 - A),
\end{equation}
which provides a general description of noise propagation through a continuum with constantly varying attenuation and temperature.
In the limits of $A\to1$, e.g., for a nearly lossless component, Eq.~\ref{eq:noise_prop_appendix} is simplified to
\begin{equation}
\begin{split}
    \mathbb{T}_{N}|_{A\to1} \approx & \mathbb{T}_{0} + \left(\frac{T_{N} + T_{0}}{2} - \mathbb{T}_{0}\right) (1 - A) \\
    = & \mathbb{T}_{0} A + \left(\frac{T_{N} + T_{0}}{2} \right)(1 - A).
    \label{eq:noise_sim}
\end{split}
\end{equation}

Eq.~\ref{eq:noise_prop_appendix} can be further generalized for a RF system with attenuation and temperature arbitrarily distributed over the body.
This can be described by the same approximation as Eq.~\ref{eq:noise_prop_condition} assuming $\delta A_n$ and $\delta T_n$ are fixed within a single segment but varying segment by segment.
Eq.~\ref{eq:noise_prop_appendix} itself describes the noise propagation within a single segment, while iteration of it develops the noise temperature with increasing segment number as follows
\begin{equation}
\begin{split}
\label{eq:noise_prop_nodes}
    \mathbb{T}_{1} \equiv \mathbb{T}_{0 \to 1} &= \mathbb{T}_{0} + (T_{1} - T_{0})\left(1 + \frac{1 - \delta A_1}{\ln \delta A_1} \right) \\
    & + (T_{0} - \mathbb{T}_{0})(1 - \delta A_{1}), \\
    \mathbb{T}_{0 \to 1 \to 2} &= \mathbb{T}_{1} + (T_{2} - T_{1})\left(1 + \frac{1 - \delta A_2}{\ln \delta A_2} \right) \\
    & + (T_{1} - \mathbb{T}_{1})(1 - \delta A_{2}), \\
    \vdots & \\
    \mathbb{T}_{0 \to 1 \to \cdots \to N} &= T_{N} + (\mathbb{T}_{0} - T_{0})A_{N} \\
    & + \sum_{n=1}^{N}(T_{n} - T_{n-1})\frac{1-\delta A_{n}}{\ln{\delta A_{n}}}\frac{A_{N}}{A_{n}}, 
\end{split}
\end{equation}
where the subscript $i \to j$ represents noise propagation from node $i$ to node $j$.
In the limit of $N\to\infty$, the continuum of a series of segments can be treated as a continuous object and then Eq.~\ref{eq:noise_prop_nodes} is rewritten in integral form of
\begin{equation}
    \mathbb{T}(s) = T(s) + \left(\mathbb{T}(0) - T(0) - \int_{0}^{s}\frac{\partial_{s^{\prime}}T(s^{\prime})}{A(s^{\prime})} ds^{\prime} \right)A(s),
    \label{eq:nois_prop_continous}
\end{equation}
where $T(s) (A(s))$ is the temperature (attenuation) at (up to) distance $s$ from the initial point.
Eq.~\ref{eq:nois_prop_continous} corresponds to the most general equation of noise propagation through a RF system with arbitrarily distributed temperature and attenuation, providing an exact estimation of noise temperature for known functions of $T(s)$ and $A(s)$.


\end{document}